\documentclass[a4paper]{article}
\usepackage{amsmath,graphicx,amssymb,mathrsfs}
\usepackage{amsfonts}
\usepackage{color}
\usepackage{subcaption}
\usepackage[margin=1in]{geometry}
\usepackage{lineno}
\usepackage{color}

\usepackage{siunitx}
\DeclareRobustCommand{\vect}[1]{
	\ifcat#1\relax
	\boldsymbol{#1}
	\else
	\mathbf{#1}
	\fi}

\usepackage{imakeidx}
\makeindex
\begin{document}
\title{Modeling Reactive Hyperemia to better understand and assess Microvascular Function: a review of techniques}	
\author{Alberto Coccarelli$^{1}$  \and Michael D. Nelson$^{2}$}
\date{$^1$ Zienkiewicz Centre for Computational Engineering, Faculty of Science and Engineering, Swansea University, UK \\
	$^2$Department of Kinesiology, University of Texas at Arlington, USA
\\corresponding author: alberto.coccarelli@swansea.ac.uk}

\maketitle

\abstract
Reactive hyperemia is a well-established technique for the non-invasive evaluation of the peripheral microcirculatory function, measured as the magnitude of limb re-perfusion after a brief period of ischemia. Despite widespread adoption by researchers and clinicians alike,  many uncertainties remain surrounding interpretation, compounded by patient-specific confounding factors (such as blood pressure or the metabolic rate of the ischemic limb).   Mathematical modeling can accelerate our understanding of the physiology underlying the reactive hyperemia response and guide in the estimation of quantities which are difficult to measure experimentally. In this work, we aim to provide a comprehensive guide for mathematical modeling techniques that can be used for describing the key phenomena involved in the reactive hyperemia response, alongside their limitations and advantages. The reported methodologies can be used for investigating specific reactive hyperemia aspects alone, or can be combined into a computational framework to be used in (pre-)clinical settings.\\


\section{Introduction}
The microcirculation is the essential ‘end point’ of the cardiovascular system, consisting of a vast network of microvessels perfusing the body’s organ tissues, whose main function is delivering oxygen and nutrients and removing waste products. At the local level, the microcirculation continuously regulates the levels of blood flow and pressure across its network to satisfy metabolic demands, to redistribute hydraulic loads and to promote inflammatory processes. To this end, a sophisticated hierarchical control system (intrinsic, metabolic and neurohormonal) operates by regulating the diameter of the microvessels, causing vaso-dilation or vaso-constriction, which in turn modulates their hydrodynamic resistance. When a significant increase in tissue metabolic demand occurs, a large fraction of the microcirculation needs to be recruited for coordinating the vaso-dilation. To achieve this, the vaso-dilation arising from the lower microcirculation can ‘ascend’ into feed arteries by means of an electrical signal conducted along the endothelial cells (ECs), hyper-polarizing vascular smooth muscle cells (SMCs) causing relaxation~\cite{bagher2011}.

Since microcirculation represents the ‘mesoscale’ functionally bridging the systemic circulation to perfused tissues, microvascular dysfunction (MVD) ultimately compromises oxygen delivery to the end organ(s)~\cite{thomas2020}. 
Indeed, it is within the microcirculation that the earliest signs of several cardiovascular diseases manifest themselves. For example, the Firefighters and Their Endothelium (FATE) study~\cite{anderson2011} showed that microvascular health represents a powerful independent predictor of cardiovascular events in primary prevention. Thus, assessment of microvascular function may not only have utility for providing novel insights into the pathophysiology of the patient, but also presents an important opportunity for early disease detection and risk stratification~\cite{albadri2019}. 

\begin{figure}[h!]
	\includegraphics[width=1\linewidth]{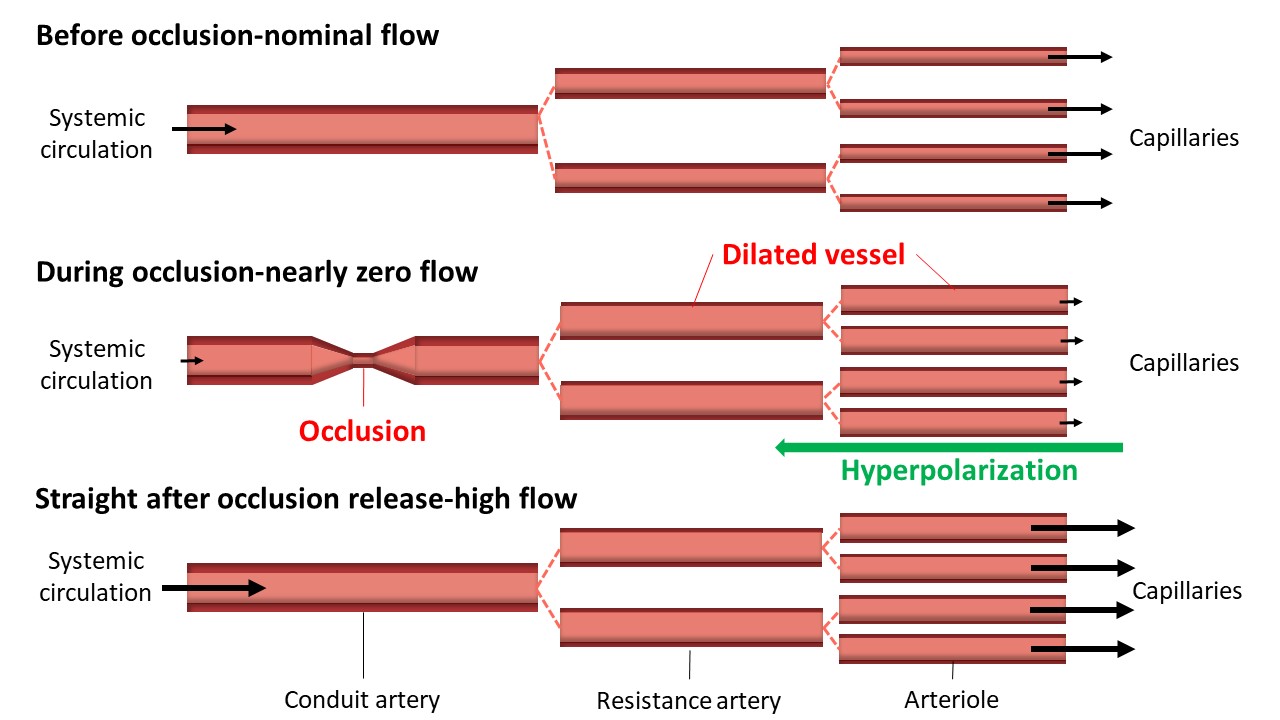}
	\caption{Microvascular conditions before, during and after occlusion. The size of the arrow indicates the magnitude of blood flow. The vasodilation in resistance arteries may have a smaller entity than in arterioles due to the limited distance that can be reached by the ascending hyperpolarizing signal.}
	\label{rh}
\end{figure}

Microvascular function can be evaluated invasively at the level of the end-organ using vaso-active agents combined with pressure/flow wires and non-invasively using advanced imaging techniques such as computed tomography (CT), magnetic resonance imaging (MRI) and positron emission tomography (PET)~\cite{camici2015,schindler2020}. Invasive approaches can indeed selectively partition macro- from microvascular function but are not suitable nor feasible in all individuals. While non-invasive CT, PET and MRI address many of these concerns, exposing participants to ionizing radiation, confined spaces, and/or strong magnetic fields also limits widespread adoption. As such, there has been considerable focus on the development of safe and cost-effective alternative methods for measuring microvascular reactivity~\cite{troy2021}.
Accordingly, much attention has been placed on the peripheral vasculature, as it is easily accessible and reasonably reflects vascular function in other end-organs of interest~\cite{albadri2019b,broxtermann2019,nardone2020}. Although not yet established in clinical practice, methods such as reactive hyperemia provide prognostically-significant indicators, capable of differentiating clinical phenotypes~\cite{rosenberry2019,rosenberry2020}. In its simplest form, reactive hyperemia represents the magnitude of limb reperfusion following a brief period of ischemia induced by arterial occlusion (see Figure \ref{rh}). 
Fundamentally, the approach measures microvascular vasoreactivity to metabolic substances produced in response to tissue ischemia.
Here we consider reactive hyperemia to be limited only to the initial increase in flow that immediately follows the restoration of flow. Indeed, the blood flow response after this initial period is affected by regulatory mechanisms that fall outside of pure `microvascular function'. Multiple methods exist for evaluating reactive hyperemia, including limb distension by venous occlusion plethysmography, blood velocity/flow via Doppler ultrasound of an upstream conduit vessel, kinetic changes in tissue oxygenation by near-infrared spectroscopy (NIRS)~\cite{huang2007,farouque2003,crecelius2013a,rosenberry2018,barstow2019}, and tissue perfusion by (near-infrared) diffuse correlation spectroscopy (NIRS-DCS)~\cite{bartlett2020}; each of which with its own strengths and weaknesses~\cite{rosenberry2020}. As detailed in this review article, the factors governing reactive hyperemia are multi-factorial and difficult to quantify. Thus, the best measurement approach is one that accounts for the greatest number of such complexities. 

Computer modeling can be used as a complementary tool for hypothesis testing, making  predictions  and  quantifying the dynamics underlying vascular regulation that cannot be directly observed during in vivo experimentation. Here, we report on various modeling techniques used to describe reactive hyperemia dynamics across the circulation, and describe the effects of various assumptions on hierarchical blood flow control system. The review concludes with an integration of knowledge, relating the theoretical modeling with existing data collected by our group and others (by way of peripheral ischemia-reperfusion), identifying key opportunities for future discovery.

\section{Haemodynamics}\label{bloodFlow}
Occlusion of a conduit blood vessel, like the brachial artery, has a direct negative impact on the resulting pressure and flow distribution downstream of the occlusion. The reduction in pressure quickly propagates from the occlusion site down to the level of the exchange vessels, ultimately impairing oxygen delivery. The ensuing tissue ischemia results in microvascular vasodilation (intended to correct the error signal), such that when the occlusion is reversed, the ensuing blood flow is markedly elevated (i.e. reactive hyperemia).

\subsection{Flow in (large) compliant vessels}
The exchange of forces between blood and vascular wall and its resulting displacement can be evaluated by employing detailed three-dimensional (3D) fluid-structure interaction models~\cite{figueroa06} or simplified one-dimensional (1D) blood flow modeling approaches which consider field variations only along the main flow (axial) direction~\cite{mynard08}. The latter methodology is less accurate (especially around localized anatomical details), but has many competitive computational advantages and can be easily adopted for vessel networks~\cite{alastruey16}. This makes 1D blood flow modeling the best option for describing the fluid mechanics in reactive hyperemia, which involves propagating phenomena along the vasculature with time scale of many seconds. Furthermore, through 1D blood flow modeling it is also possible to perform wave intensity analysis which allows the quantification of pressure waveforms travelling forward to the microcirculation and backward to the heart~\cite{mynard08}.
In arteries (and arterioles) the blood behaviour can be generally approximated as homogeneous and Newtonian since the size of the red blood cells carried by the plasma is considerably smaller (10 times) than the vessel diameter. Furthermore, flow is also generally considered incompressible and laminar, with a Poiseuille velocity profile. 
In 1D blood flow modeling, each compliant vessel can be treated as axisymmetric, with blood velocity ($u$), pressure ($P$) and flow described as continuous variables along its axial direction $z$. These quantities are averaged over the cross-sectional area ($A$) and their variation along the radial direction is considered negligible. The Navier-Stokes equations for 1D blood flow in compliant vessels can be expressed in terms of cross-sectional area and velocity averaged over the cross-section:
\begin{equation}\label{sys}
	\begin{split}
		\frac{\partial A}{\partial t}+\frac{\partial (A u)}{\partial z} = 0,  \\
		\frac{\partial u}{\partial t}+u\frac{\partial u}{\partial z}+\frac{1}{ \rho} \frac{\partial P}{\partial z} +\frac{\mu u}{\rho}\frac{8\pi}{ A} = 0,
	\end{split}
\end{equation}
where $t$ is time, $\mu$ is the fluid dynamic viscosity, $\rho$ is the fluid density, while $Q=Au$ is the (volumetric) flow rate. It is worth noting that (\ref{sys}) can also be written in terms of flow rate and pressure~\cite{carson17} or cross-sectional area and flow rate~\cite{sherwin03}.  
The mechanics underlying the vascular wall deformation appears to be complex, mainly due to vessel visco-elastic properties and the capacity to produce active tone for diameter regulation. To describe the interaction between blood and vessel wall, different approaches can be used~\cite{sherwin2003,mynard2015b,coccarelli2021a}. In the simplest case, the fluid pressure is related to the cross-section of the vessel via a linear function with respect to the luminal diameter ($D$=$\sqrt{4A/\pi}$)
\begin{equation}\label{FSI1}
	P=P_{ext}+\beta (\sqrt{A}-\sqrt{A_0}),
\end{equation}
where $P_{ext}$ is the external pressure from the surrounding tissue, $\beta$ is a parameter representing the wall elasticity and $A_0$ is the unstressed cross-section area. However, modeling reactive hyperemia requires a vessel wall model able to describe the hyperpolarization-induced dilation and then the resulting (compliant) structural response to hyperemic flow. Given such complexity, wall mechanics models derived from conservation laws retaining mechanobiological features~\cite{murtada2014,coccarelli2018} are preferable over tube laws which are purely phenomenological; indeed the latter, for this specific application, would require the introduction of several non-physical parameters. It is also noted that, if desired, wall viscoelasticity can be integrated into the blood pressure-wall deformation law by using more complex constitutive models~\cite{alastruey2011b,perdikaris2014,bertaglia2020} with a consequent decrease in computational efficiency. The haemodynamic features (viscosity and density) and the geometric (diameter and length) and structural (stiffness) wall properties can be made vessel specific and can reflect different physiological and pathological conditions such as ageing or hypertension~\cite{mynard08,coccarelli2018b}. 
The blood flow variables described by system (\ref{sys}) and (\ref{FSI1}) can be computed by employing a broad variety of numerical schemes including finite differences, finite elements and finite volumes and can be extended to large vessel networks by imposing mass and momentum conservation at the interface between vessels~\cite{stergiopulos1992,mynard08,muller2013,carson17}.

\subsection{Biphasic flow in microvessels}
\begin{figure}[h!]
	\includegraphics[width=1\linewidth]{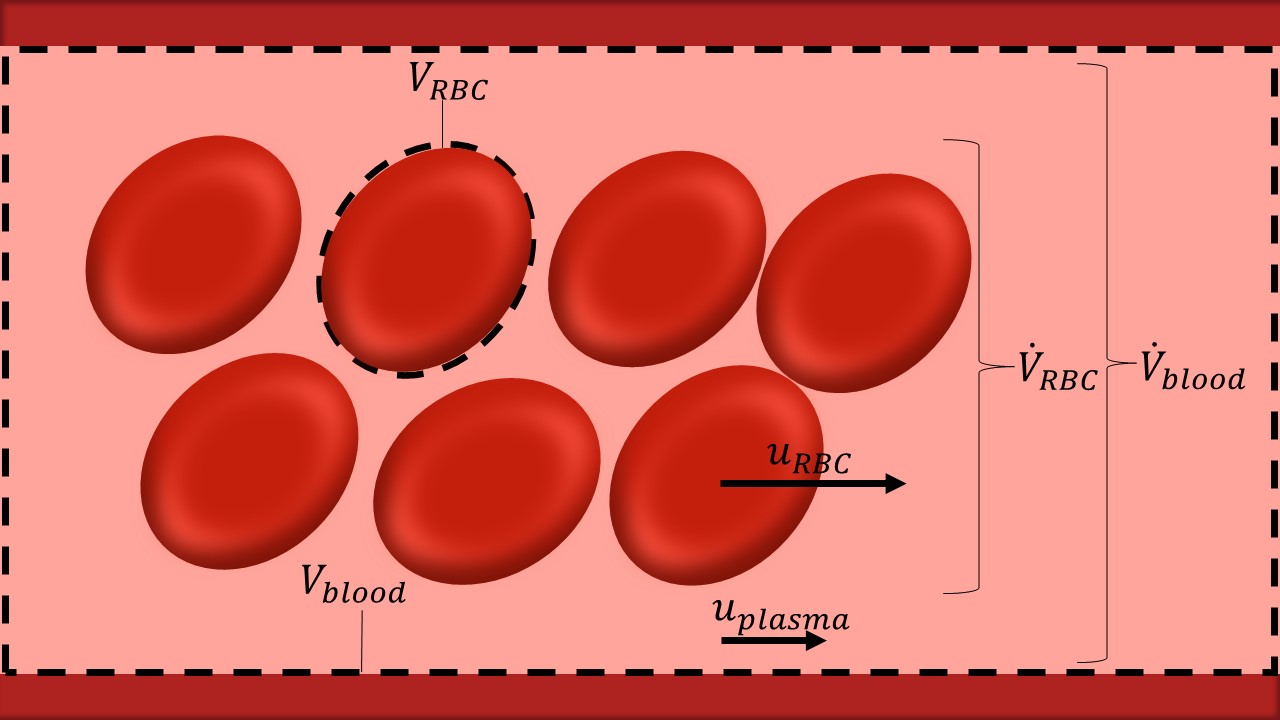}
	\caption{Hematocrit in blood vessel. The discharge hematocrit, being the volume fraction of RBCs in blood delivered through the vessel by the flow, is obtained as ratio the vessel volumetric rate of RBCs and volumetric blood flow ($H_D$=$\dot{V}_{RBC}/\dot{V}_{blood}$). The tube hematocrit is given as ration between the volume occupied by RBCs and total blood volume in the vessel ($H_T$=$n_{RBC}{V}_{RBC}/{V}_{blood}$, where $n_{RBC}$ is the number of RBCs in the vessel).}
	\label{hematocrit}
\end{figure}

During reactive hyperemia, flow in the microvessels is altered from its physiological range due first to the sudden pressure reduction induced by the upstream occlusion and then by arteriolar luminal expansion consequent to the wall relaxation driven by ischemic tissues.
Due to its morphology and function, the downstream vasculature constitutes the site of major blood pressure drop along the cardiovascular system. The work by Secomb~\cite{secomb2017} provides a detailed characterization of the flow through microcirculatory networks.
In these microvessels Reynolds number is $<$ 1, and therefore the blood flow can be described with a good level of accuracy as incompressible Stokes fluid, for which the convective component is neglected. Blood flow can still be described by (\ref{sys}) but most assume a rigid microvessel wall, which implies considering only the momentum conservation equation for relating blood flow and pressure.
Furthermore, the biphasic nature of flow requires a modeling framework that accounts for the main rheological properties of its components. 
Since they are concentrated in the central part of the tube, RBCs travel faster than the surrounding plasma (referred as Fahraeus effect) and the shorter RBC's transit time with respect to the total blood flow has a profound effect on oxygen transport and exchange with tissues. The discharge hematocrit ($H_D$), which is the ratio between RBCs volume flux and total blood volume flux along the tube, appears higher than the tube hematocrit ($H_T$), which is the volume occupied by RBCs over the tube volume at a certain time instant (see Figure \ref{hematocrit}). 
The mean velocity of RBCs ($u_{RBC}$) can be estimated if the ratio between $H_T$ and $H_D$ is known (since ${H_T}/{H_D}$=$u/{u_{RBC}}$), for which Pries et al.~\cite{pries1990} identified the following correlation:
\begin{equation}
	\frac{H_T}{{H}_D}=H_D+(1-H_D)(1+1.7 \text{exp}(-0.35D)-0.6 \text{exp}(-0.01D)).
\end{equation}
The latter expression was then modified by Pries et al.~\cite{pries2005b} to account for the reduction in diameter due to the endothelial surface layer.
In tubes with diameters below 300 $\mu$m, the viscosity of blood reduces with decreasing diameter. This phenomenon, discovered by Fahraeus and Lindquist~\cite{fahraeus1931} (referred as Fahraeus-Lindquist effect) appears to be associated with the tendency of RBCs to migrate away from the vessel wall~\cite{secomb2017}. The resulting blood viscosity can be evaluated as a product between the plasma viscosity and the relative apparent viscosity ($\mu_{rel}$), with the latter accounting for the effect of the hematocrit. Pries and co-workers defined a set of correlations which relate apparent viscosity with discharge hematocrit, and diameter of microvessel for in-vitro~\cite{pries1992} and in-vivo conditions~\cite{pries1990,pries1994}. In the latter case, the relative apparent viscosity is defined as:
\begin{equation}
 \mu_{rel}=[1+(\mu_{45})\frac{(1-H_D)^{C_{\mu}}-1}{(1-0.45)^{C_{\mu}}-1}(\frac{D}{D-1.1})^2](\frac{D}{D-1.1})^2,
\end{equation}
in which $\mu_{45}$ and $C_{\mu}$ are experimentally derived correlations dependent on the vessel diameter.
The presence of cell-free layers within the microvessels is also responsible for the phase separation effect, which consists in a non-proportional partition of RBCs and plasma flows at diverging bifurcations.
The experimentally-derived correlations by Pries et al.~\cite{pries1989,pries2005b} represent the reference approach for including this phenomenon within vascular networks.
An alternative RBCs phase separation model was subsequently introduced by Gould and Linninger~\cite{gould2015}, although its benefits over the one by Pries et al.~\cite{pries1989,pries2005b} have been debated~\cite{rasmussen2018}.

Opposite to the continuum approach, blood flow across microvessels can also be described via motion of multiple discrete particles~\cite{gambaruto2015,fedosov2010b} and the effect associated to the particulate nature can be assessed for different channel sizes~\cite{bagchi2007}. 
To transit across the smallest capillaries, the RBC membrane undergoes significant deformation, with the intensity of applied forces on the capillary lumen playing a significant mechanobiological role~\cite{harraz2022}. We expect that the surge in blood flow during reactive hyperemia substantially changes the RBC-capillary lumen interaction. Variation in RBC number and velocity across the capillary bed can be experimentally analyzed by using intravital video recordings~\cite{mendelson2021} but the resulting pressure field remains uncertain.
Discrete multi-scale RBC modeling approaches~\cite{fedosov2010a,chang2016,ademiloye2018} can be used for reproducing the mechanical and rheological properties of the cell and their interaction with plasma. 
Casquero et al.~\cite{casquero2017} used a NURBS-based fluid-structure interaction framework for assessing the deformation of RBCs, which were modeled as compound capsules (or solid membrane) where the nucleus is treated as bulky deformable solid.
These multidimensional models allow the direct assessment of the physics underlying the interaction between RBCs and microvessel wall but they require efficient parallel algorithms~\cite{grinberg2011,perdikaris2016} for being scalable to realistic and physiologically relevant problems. 



\subsection{Lumped circulation models}
Lumped or zero-dimensional (0D) modeling represents an approach in which the system's quantities are not (directly) space-dependent but depend only on time. Their governing equations can be directly derived from formulations of high-fidelity/space-dependent models (such as full 3D heart chambers, arteries and veins) and the space-dependent characteristics are encapsulated through a set of parameters which are generally optimized using experimental data~\cite{stergiopulos1995}.
0D models may be employed alone for describing the full cardiovascular system or parts of it and can be represented through electrical analogy.
Lumped models can also be used in combination with higher-fidelity counterparts to represent those vascular territories where a low level of space details is required.

Compartmental lumped models~\cite{braakman1989,olufsen2005,mohammadyari2021} split the vasculature into different flow compartments, each of them characterized by specific parameters representing vascular features such as flow resistance and compliance. In this case, the full haemodynamic profiles across the vasculature can be obtained by simply computing a set of (time-dependent) ordinary differential equations. This approach involves very low computational costs and allows an easy integration and/or comparison with quantities averaged over space obtained from some experimental imaging techniques (such as doppler ultrasound or NIRS-DCS).
Therefore, due to its simplicity, 0D modeling should be considered as first-step for modeling blood flow variables during reactive hyperemia.
On the other hand, lumped modeling alone cannot adequately capture the wave-propagation phenomena underlying the reactive hyperemia response and therefore does not provide a suitable platform for a full mechanistic understanding of the different space-dependent interactions between vascular territories. Nevertheless we re-iterate that, for various questions related to reactive hyperemia, lumped modeling should be considered for the description of the whole system's dynamics, with space-dependent models introduced as `sub-models' for further refinement if necessary.

\subsection{Vascular network definition}
The topology of the fluid network can either be reconstructed from images~\cite{boileau2017,smith2014,mendelson2021} or generated by using branching (either deterministic or stochastic) algorithms~\cite{adjoua2019,koppl2020}. 
From CT or MRI scans it is possible to obtain, after segmentation, networks of large blood vessels~\cite{boileau2017} but diameters smaller than 500 $\mu$m are generally below the standard angiography spatial resolution. On the other hand, microvascular networks can be digitally reconstructed, down to the capillary scale, by leveraging recent developments in high-resolution microscopy~\cite{smith2014}. 
To facilitate the integration of experimental datasets in network models, methods were developed for classifying microvessels without requiring any flow information~\cite{smith2014} or estimating flow rate with unknown boundary conditions~\cite{fry2012}. 
The microvascular arrangement and consequent levels of blood perfusion strongly depend on local morphological and functional properties of the tissue such as O$_2$ consumption rate, level of venous drainage and presence of anastomosis~\cite{zeller2017}. Skeletal muscles are perfused by bundles of parallel capillaries which transport blood from a common terminal arteriole to the same post-capillary venule. Mendelson et al.~\cite{mendelson2021} used intravital video microscopy to show that capillary modules may present substantial structural differences in terms of topology, as well as functional heterogeneities in RBCs haemodynamics within the same module and across modules. This study proposed a new paradigm for capillary flow organization which is pivotal for explaining the complex flow regulation occurring during skeletal muscle contraction. Capillary modules are indeed interconnected to form continuous columns, defined as capillary fascicle, that align with the structure of the muscle fascicle~\cite{mendelson2021}. Blood flow through the arterial-venous vasculature (arteries $\rightarrow$ arterioles $\rightarrow$ capillaries $\rightarrow$ venules$\rightarrow$ veins) is complex, dictated by varying vessel wall morphology and geometry. Due to this, multi-scale network models can be used to accurately capture the blood flow transition across the mesoscale (large to microvessel)~\cite{adjoua2019,koppl2020}.

To compute haemodynamic variables along a portion of the vasculature, it is necessary to define the boundary conditions at the network extremities to represent the remaining part of the cardiovascular system that is not explicitly modelled in space (see for instance~\cite{alastruey2007}). For computing blood flow in large arteries during reactive hyperemia, the inlet boundary condition could be imposed either at the heart level or at a (proximal) brachial artery site (as long as the inlet is not beyond the occlusion site). However, the latter case is practically less feasible than the former as it requires the prescription of both flow and pressure (which are difficult to measure simultaneously) for not omitting the potential contribution of backwards reflections. The left ventricle (or the proximal aortic node) can therefore be used as the inflow prescription point~\cite{sherwin2003,formaggia2006,mynard08,blanco2015}, with its complex pumping function simplified using a lumped model, which is able to capture the essential features of inflow data. 

To close the fluid network, it is always necessary to define appropriate outflow conditions, which represent the haemodynamics characterizing the downstream systems.
Venous hydrodynamic conditions (i.e., position of the limb) may play a significant role in the experimental settings~\cite{bartlett2020}, as they directly affect the downstream pressure level. {Therefore. inclusion of a part of the venous system into the computer model may be useful for reflecting changes in the functional state of its components}. 
Others have also proposed using a closed loop flow model, in which the arterial network system is coupled with a venous counterpart at the heart and capillary levels~\cite{muller2014,mynard2015b}. While this holistic approach incorporates additional features, like the venous network, translation of this model to a patient-specific context is hindered by the large numbers of parameters which need to be identified/measured. Due to their simplicity and adaptability, lumped Windkessel models represent the most popular approach for modeling vascular territories beyond the main system of interest~\cite{mynard2015b}. These lumped models can be combined in series to form complex interconnected compartments, allowing to establish a link between their functional states and the outflow conditions.
Alternatively, more complex methods can be used to represent the downstream circulation, including the structured (or fractal) tree approach~\cite{olufsen1999}, its derived versions~\cite{cousins2012,perdikaris2015}, and the porous-media based approach~\cite{coccarelli2019,coccarelli2021b}. While these models encapsulate more details of the downstream network, they come with higher computational costs.

\section{Coordinated regulation}\label{control}
Tissue ischemia generates hyperpolarization of the smallest microvessels (capillaries mainly) which is then conducted upstream through EC gap junctions to relax the arteriolar vasculature and favour a more consistent/systemic increase in blood flow. In the time period between tissue ischemia generation and the full recovery of the system, intrinsic pressure and flow control mechanisms simultaneously work together to restore homeostasis; the entirety of which is extremely complex and difficult to quantify~\cite{carlsson1987,koller2004}. 

\subsection{Generation of the hyperpolarizing signal}
Tissue metabolism involves the production and consumption of different substances, some of which have been traditionally associated with hypoxia (i.e., adenosine, K$^+$, CO$_2$, lactate and H$^+$).
In reactive hyperemia, the mismatch between oxygen demand and supply is caused by a sudden decrease in oxygenated blood flow and not by a change in tissue metabolic activity, which instead drives functional hyperemia.
The review by Rosenberry and Nelson~\cite{rosenberry2020} reports an exhaustive description of the cellular pathways involved from the stimulus generation to vessel dilation. Accordingly, the stimulus originating the hyperpolarization is yet to be defined but most likely involves a combination of K$^+$, ATP, bradykin, H$_2$O$_2$ and epoxyeicosatrienoic acids. These molecules cause vessel hyperpolarization by activating different vascular channels including inwardly rectifying potassium channels and Na$^+$/K$^+$ pumps~\cite{crecelius2013a}. Recent findings~\cite{longden2021} on the capillary EC function showed that membrane hyperpolarization may trigger a complex cascade of Ca$^{2+}$ signaling pathways which include cross-talk between intracellular Ca$^{2+}$ stores and transient receptor potential channels.
In a hypoxic environment, RBCs may also contribute to microvascular dilation by releasing substances into the bloodstream such as ATP~\cite{ellsworth2012,sprague2012,ellsworth2016,keller2017,zhou2019,ghonaim2021}.

In the computer model, the originating stimulus can be represented either as a single variable in a set of kinetic equations describing oxygen/tissue signaling or simply as perturbation of the concentration of species in the extracellular space. In the reference model by Moshkforoush et al.~\cite{moshkforoush2020} the effect of cerebral metabolic activity on capillary ECs was represented as an increase in extracellular $K^+$ concentration, which in turn affects transmembrane K$^+$ channels and membrane potential.
Hyperpolarization of capillaries can be modeled by employing electrochemical models~\cite{wiesner1996,silva2007,moshkforoush2020,coccarelli2022a} describing the dynamics of intracellular species (such as Ca$^{2+}$, K$^+$ and Na$^+$) and membrane potential in stimulated ECs. 
These models are considered as `space-homogeneous` (or lumped/0D) because their variables represent quantities averaged over the cell volume and therefore only dependent on time. In the seminal work by Wiesner et al.~\cite{wiesner1996} the effect of an exogenous agent (thrombin) was simulated by combining the EC Ca$^{2+}$ dynamics with kinetics models representing the time evolution of ligands involved in the receptor-activation pathway.  
Modeling studies that quantitatively characterize the function of a component of this orchestra, such as~\cite{coccarelli2022a}, are therefore instrumental for deciphering the emergent cell dynamics features.
The capabilities of space-homogeneous models can be further extended by adding transport equations for describing diffusion of species~\cite{wiesner1997} or models representing cytosolic micro-domains~\cite{moshkforoush2019} for assessing the interaction between sub-cellular components.
Space-homogenous models allow an easy extension to cluster of cells for computing the resulting intra-cellular signaling. On the other hand, this would be hard to achieve in case a multi-dimensional approach for cell dynamics was used. Therefore, cellular space-homogeneous models represent the most trivial option as bottom level of a multi-scale vessel framework.
Furthermore, O$_2$-induced release of ATP from RBCs across capillary networks can also be accounted for in the model by following the work by Goldman et al.~\cite{goldman2012,ghonaim2013}.
However, a recent theoretical study~\cite{reglin2017} suggested that, with respect to other O$_2$-driven regulatory pathways, the role of vasoactive molecules release by RBCs is limited.

\subsection{Conducted vasodilation}
Ascending vasodilation allows to export the local capillary hyperpolarization to a widespread fraction of the microvasculature. The endothelium serves as the predominant cellular pathway for signal conduction as the hyperpolarization is transmitted from cell to cell along the vessel wall through gap junctions~\cite{bagher2011}. 
According to Welsh et al., the conduction of the signal is enabled by a defined pattern of charge movement along the vascular wall which is the result of the interplay between tissue structure, gap junction resistivity and ion channel activity~\cite{welsh2018}.
Despite the vessel axis being the main direction of propagation, the hyperpolarizing signal is also transmitted into surrounding SMCs through myoendothelial junctions to promote vascular relaxation~\cite{tran2012}.
Endothelial Ca$^{2+}$-activated K$^+$ (SKCa/IKCa) channels seem to play a key role in tuning electrical conduction along microvessels by modulating signal dissipation through changes in transmembrane resistance~\cite{behringer2012}.
Experiments in rat cremaster arterioles indicate that vasoactive conduction is favoured by circulation of autocrine and paracrine mediators ATP and K$^+$, which is increased under hyperemic conditions~\cite{dora2017}.

Early efforts in modeling electrical communication across vascular branching include the works by Diep et al.~\cite{diep2005}, in which vascular intercellular gap junctions in the skeletal muscle were represented by ohmic resistors.
For investigating the signal transduction involved in conducted vaso-reactivity, Kapela et al. integrated their previous modeling efforts~\cite{silva2007,kapela2009} into a multicellular model of a rat mesenteric arteriole which comprises a detailed description of ECs-SMCs units, coupled via non-selective gap junctions~\cite{kapela2010}. This framework was used in a subsequent study~\cite{kapela2018} for estimating from experimental measurements the electrical resistance associated to cell membrane and gap junctions.
More recently, the same group presented an experimentally-validated computational framework for reproducing the propagation of the hyperpolarizing signal across a cerebral microvascular network~\cite{moshkforoush2020}. This provides some key information regarding how the activity of different ion channels can impact on the ascending signal and its re-generation. However, this framework neither quantifies the dynamic changes in microvascular compliance nor indicates whether the metabolic demand is satisfied but it represents a solid modeling base to start from.
Alternatively, Arciero et al.~\cite{arciero2008} developed a multi-compartmental framework based on a previous autoregulation wall mechanics model to represent the adjustments in diameter across a microvascular network induced by vaso-conducted metabolic response. In this work the stimulation considered was the hypoxic-induced ATP release by RBCs and the hemodynamic model was coupled with Krogh-type models for representing the mass exchange with tissue.

\subsection{Vessel dilation}
The conducted changes in membrane potential spread from capillaries to arterioles. Indeed, in rabbit skeletal muscle~\cite{bosman1995,bosman1996} and studies on brain function~\cite{hall2014} capillaries are able to control local blood through pericytes contraction.
In the study by Horn et al.~\cite{horn2022}, a feed artery occlusion caused a heterogeneous flow distribution in capillary beds with the presence of capillary no-reflow, and this appeared as result of the complex interaction between myogenic and metabolic mechanisms occurring at the arteriolar level. 
In simulated capillary modules, RBCs flow appeared to be independent from the network resistance~\cite{mendelson2022}. This indicates that RBCs flow across capillaries is mainly regulated at control sites located at-pre and post-capillary levels.
Overall, the role of single capillaries on skeletal muscle local flow regulation still needs to be fully elucidate but its extent seems minor with respect to the upstream microcirculation~\cite{poole2020}. Experimental evidence indicated that also venules and veins are endowed with a contractile apparatus, and therefore they may play a role in the pressure re-distribution occurring during and after the arterial occlusion~\cite{davis1988,szentivanyi1997,dongaonkar2012}. However, their high blood volume capacity suggests a minor role of these vessels especially in the first part of haemodynamic response after occlusion release.

The level of dilation of each microvessel depends on the magnitude of the hyperpolarization signal that has reached its location and by the ion channels expression levels of its wall constituents~\cite{murtada2018}. A change in the wall membrane potential promotes the increase in vessel diameter in potentially two ways. First, as direct effect, hyperpolarization of the SMC causes a decrease in intracellular Ca$^{2+}$ which ultimately diminishes the activity of myosin light chain kinase with reduction of cross-bridges formation. A second effect on the contractile machinery may be mediated through ECs' Ca$^{2+}$ dynamics. Indeed, in these cells, the decrease in membrane potential causes an increase in intracellular Ca$^{2+}$ which promotes pathways the formation of autocoids such as NO which ultimately affect the cross-bridges by phosphorylating the myosin light chain phosphatase. 
Both pathways, by reducing the fraction of attached actin and myosin filaments, leads to distension of the SMC and consequent vessel dilation. However, Crecelius et al.~\cite{crecelius2013a} showed that, in the forearm, there is no combined role of NO and PGs in peak RH. Therefore, EC Ca$^{2+}$ dynamics appears to be crucial for the conduction of the hyperpolarizing signal along the endothelium and to the adjacent SMCs while its paracrine function is less relevant. In skeletal muscle resistance arteries, Ca$^{2+}$ sensitization pathways are also known to play a role in the myogenic response~\cite{moreno2013}, but these seem to be present only above a certain blood pressure level ($\sim$ 60 mmHg)~\cite{osol2002}.

To accurately model the change in vessel diameter due to membrane hyperpolarization a mechanistic multi-scale model is required. The contractile machinery model needs to translate the biochemical information from SMC Ca$^{2+}$ signaling into tissue deformation.
For this purpose space-homogenous models for describing cellular signaling seems the most viable choice. The SMC Ca$^{2+}$ dynamics model developed by Kapela et al.~\cite{kapela2008}, based on 26 partial differential equations, provides a comprehensive description of the interaction between several ionic channels upon different agonist stimulation.
Pioneering work by Yang et al.~\cite{yang2003a,yang2003b} bridges the multi-scale components of SMC contractility. This framework, devised for computing the myogenic response in rat cerebrovascular arteries, includes a description for the cross-bridges kinetics and corresponding cell length variation, which establishes a mechanistic link between intracellular Ca$^{2+}$ variation to structural vessel deformation. Similarly, Layton's group developed other reference models~\cite{chen2011,edwards2014} able to accurately mimick the myogenic response in renal afferent arterioles.
On the contrary, Carlson and Secomb~\cite{carlson2005} presented a minimalistic vessel wall mechanics model that, without including any cellular/sub-cellular component, provides a good agreement for both passive and active tension against myograph experimental data from single isolated microvessels. This type of approach is extremely attractive due to its simplicity and limited computational cost but a dynamic interplay with ECs network carrying the hyperpolarizing signal remains challenging to establish. 
The multi-scale models following the approach by Murtada et al.~\cite{murtada2010,murtada2012,murtada2014,coccarelli2018} provide a rigorous way to encapsulate the information regarding the actin-myosin interaction into a tissue/continuum mechanics level by using an active strain energy function. This class of models represents the best option for coupling the resulting wall deformation and stress with the blood flow within the vessel.

\subsection{Regulated networks and clinical data integration}
To date, there is a limited number of models able to describe the coordinated haemodynamic response to arterial occlusion through a multi-scale vascular system. 
The work by Yamazaki and Kamiyama~\cite{yamazaki2014} is one of the earliest attempt to simulate flow-mediated dilation in a conduit vessel by including multiple scales (from cell to vessel) into one modeling framework. Jin et al.~\cite{jin2020}, extended this work by including a realistic blood flow network model. This model closely approximates experimental recordings made at the conduit artery level but its representation of the downstream microvascular dynamics remains limited. Some studies~\cite{schmid2015,schmid2017,gkontra2019} have also assessed the impact of microvascular diameter changes on blood rheology and tissue perfusion across different vasculatures. It is however important to note that, despite such models allowing quantification of flow characteristics upon an imposed variations in microvessel resistance, there is no direct link with a realistic coordinated regulatory response.

\begin{figure}[h!]
	\includegraphics[width=1\linewidth]{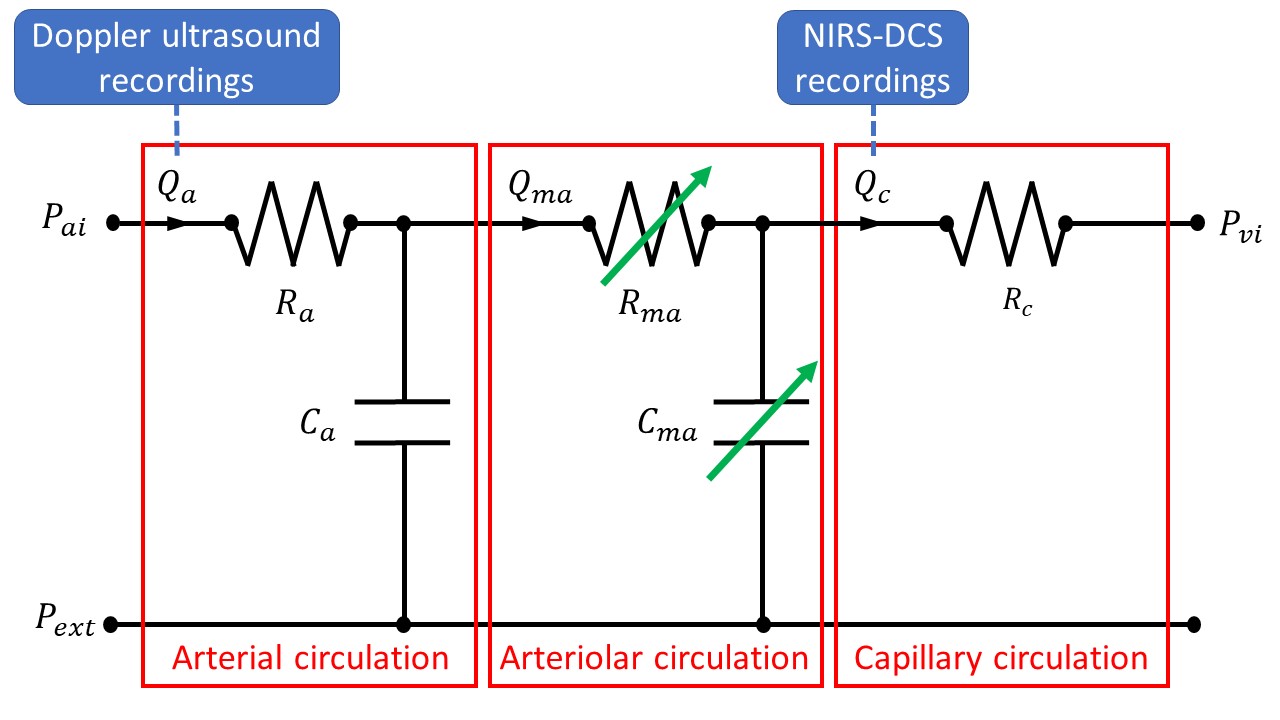}
	\caption{Example of regulated blood flow compartmental model with potential clinical data integration. $P_{ai}$: inlet arterial pressure; $Q_{a}$: arterial flow; $R_{a}$: arterial resistance; $C_{a}$: arterial compliance; $Q_{ma}$: micro-arterial/arteriolar flow; $R_{ma}$: micro-arterial/arteriolar resistance; $C_{ma}$: micro-arterial/arteriolar compliance; $Q_{c}$: capillary flow; $R_{c}$: capillary resistance; $P_{vi}$: inlet venous pressure; $P_{ext}$: external pressure from surrounding tissue. $R_{ma}$ and $C_{ma}$ are modulated to reflect vascular changes due to regulatory mechanisms (green arrows).}
	\label{circ}
\end{figure}

Due to their intrinsic simplicity, compartmental lumped models represent the standard approach for evaluating the blood flow dynamics across different levels of the micro-vasculature, without directly describing propagation phenomena. For each compartment, vascular changes due to regulation can be represented by varying (in time) its lumped parameters (such as resistor and capacitor, see Figure \ref{circ}) according to a more or less complex internal variable evolution equations~\cite{demul2005,vo2007,solovyev2013,chen2017,zhaoE2020}.
de Mul et al.~\cite{demul2005,demul2009} were among the first to describe the fluid-dynamics occurring during post-arterial occlusion reactive hyperemia by using compartmental modeling, through which vascular territories are split into arterial and capillary systems. In~\cite{demul2005}, measurements from Laser-Doppler perfusion monitoring were used for estimating two time constants characterizing the dynamics, which differed across different types of patients (healthy vs unhealthy). Likewise, Vo et al.~\cite{vo2007} modelled reactive hyperemia response to venous occlusion and by using NIRS data.
Solovyev et al.~\cite{solovyev2013} defined an hybrid model, combining a lumped model for blood flow with an agent based model of skin injury, to assess the role of blood flow on the development of pressure ulcers in spinal-cord injury patients. Recently, a multi-compartmental model incorporating oxygen transport, tissue metabolism, and vascular regulation mechanisms was proposed by Chen and Wright~\cite{chen2017} to characterize and interpret MRI-derived microvascular reactive hyperemia (by arterial spin labeling) in calf muscles of human volunteers.
Through all these modeling approaches it is generally possible to evaluate quantities that can be directly associated with in-vivo experimental measurements carried out in human patients. 
However, this type of approach requires the condensation of most regulatory mechanisms within few model parameters, which might not have a clear physical meaning and that in most of the times need to be estimated by experimental fitting. At the same time, some of these compacted model parameters can be used for a straightforward discrimination of different types of cardiovascular profiles. As such, given the complexity of reactive hyperemia, this class of models is likely most appropriate for capturing the {interaction between vascular compartments and their contribution to} the regulated haemodynamic response.

\section{Discussion and concluding remarks}\label{disc}
There is a pressing need to enhance our understanding on microvascular function and its regulation. Advances in this direction will ultimately lead to the development of new diagnostic, progress monitoring and therapeutical approaches for diseases characterized by (micro-)vascular dysfunction.
Through this work, we sought to provide a guide for modeling reactive hyperemia, which represents a standard method for assessing microvascular reactivity.
Based on the current knowledge on reactive hyperemia, we reported a series of models that can be used for describing physiological phenomena that occurs at different space and time scales, spanning from cellular ionic currents to systemic circulation flow dynamics.
Several issues hinder the accurate modeling of reactive hyperemia. First and foremost, from an experimental point of view, isolating single vascular control mechanisms and vascular compartment contribution to the resulting haemodynamics is technically very challenging if not impossible. On the other hand, it is possible to measure the combined response at locations across different experimental conditions, in order to best inform the computer model.
The use of multiple simultaneous recordings of blood flow via Doppler ultrasound as well as RBCs flux via NIRS-DCS may also be helpful in the {identification of model parameters and quantification of their uncertainties}~\cite{englund2013}.

There is need to define, test and use suitable modeling approaches that, despite the complexity of the system analysed, are capable of capturing the relevant interactions between vascular compartments. Furthermore, there is need for computational models that can efficiently deal with different scales in space and time as this aspect is essential for the successful translation of the computer model into clinical practice.
Connections between components of the hierarchical multi-scale models play a crucial role. Further investigation and validation is needed before modeling may inform clinical decision making. 
To this end, high-fidelity models that leverage large existing datasets augmented by machine learning/data-driven approaches may be needed.


\section*{Declaration of competing interest}
The authors declare that there is no known competing financial interests or personal relationships that could have appeared to influence the work reported in this paper.

\bibliography{ACreferences}
\bibliographystyle{unsrt}


\end{document}